\documentstyle[preprint,prc,aps,psfig]{revtex}

\tightenlines

\begin{document}

\title{Generator Coordinate Truncations} 
\author{K. Hagino,$^{1,3}$ G.F. Bertsch,$^{2,3}$ and 
P.-G. Reinhard$^{4}$}
\address{$^1$Yukawa Institute for Theoretical Physics, Kyoto
University, Kyoto 606-8502, Japan }
\address{$^2$
Institute for Nuclear Theory and Department of Physics, \\
University of Washington, Seattle, WA 98195}
\address{$^3$Institut de Physique Nucl\'eaire, IN2P3-CNRS, \\ 
Universit\'e Paris-Sud, F-91406 Orsay Cedex, France}
\address{$^4$
Institut f\"ur Theoretische Physik II, Universit\"at Erlangen-N\"urnberg, \\
Staudtstrasse 7, D-91058 Erlangen, Germany} 

\maketitle

\begin{abstract}

We investigate the accuracy of several schemes to calculate ground-state 
correlation energies using the generator coordinate technique. 
Our test-bed for the study is the $sd$ interacting boson model, equivalent 
to a 6-level Lipkin-type model. 
We find that the simplified projection of a triaxial 
generator coordinate state using the $S_3$ subgroup of the rotation 
group is not very accurate in the parameter space of the Hamiltonian of 
interest. 
On the other hand, 
a full rotational projection of an axial generator coordinate state 
gives remarkable accuracy. 
We also discuss the validity of the simplified treatment 
using the extended Gaussian overlap approximation (top-GOA), 
and show that it works reasonably well when the number of boson 
is four or larger. 

\end{abstract}


\section{INTRODUCTION}

The self-consistent mean-field theories with a phenomenological
nucleon-nucleon interaction 
have enjoyed a success in describing ground state properties
of a wide range of atomic nuclei with only a few adjustable parameters 
(see Ref.\cite{BHR03} for a recent review). 
They are now at a stage where the ground state correlations 
beyond the mean-field approximation have to be taken into account seriously. 
This is partly due to the fact that 
much more accurate calculations have been increasingly required in 
recent years 
because of the experimental progress in 
the production of nuclei far from the stability line, where the ground state 
correlation beyond the mean-field approximation may play an important 
role. 
The major part of the correlations produces effects which have
smooth trends with proton and neutron number. These are already
incorporated into the energy functionals of effective mean-field
models as, e.g., Skyrme-Hartree-Fock or the relativistic mean-field
model. However, the correlations associated with low-energy modes show strong
variations with shell structure, and cannot be contained in a smooth
energy-density functional. This concerns the low-energy quadrupole
vibrations and all zero-energy modes associated with symmetry
restoration. In fact, 
the correlation effects appear most dramatically for these
symmetry modes as there are:
the center of mass localization, the rotational symmetry, and 
the particle number conservation. Those correlation effects must be 
taken into account explicietly in order to develop a global theory which 
can be extrapolated to the drip-line regions.

There are many ways that correlation energies can be calculated.  
In Ref. \cite{HB00}, we investigated a method which uses the 
random phase approximation (RPA). We found that the RPA provides 
a useful correlation around a spherical as well as for a well deformed 
configurations, but it fails badly around 
the phase transition point between spherical and deformed. 
Because of this defect, the RPA approach does not seem the best method 
for a global theory. 
Recently, we have developed an alternative method, called the top-GOA, 
to calculate the 
ground state correlation energies based on the generator coordinate 
method \cite{HRB02}. This is a generalization of the Gaussian Overlap 
Approximation (GOA) by taking into account properly the topology of the 
generator coordinate\cite{R78}. This method can be easily applied to 
the variation after projection (VAP) scheme, 
where the energy is minimized after 
the mean-field wave function is projected on to the eigenstates of the 
symmetry \cite{RS80}. We have tested this method on the three-level 
Lipkin model, which consists of one vibrational degree of freedom and 
one rotational \cite{HRB02}, and have confirmed that the method provides an 
efficient computational means 
to calculate ground state correlation energies for the full range of
coupling strengths.

In this paper, we continue our study on the 
correlation energies 
using a model which contains the full degrees of freedom of quadrupole
motion.
To this end, we use a $sd$ interacting boson 
model (IBM)\cite{IBM,GK80}, which may be viewed as a 6-level extension of the 
Lipkin model \cite{LMG65}. 
The IBM is particularly tailored for the
description of the low-lying collective modes, thus provides a good testing
ground for the present studies of correlations. 
In realistic systems, treating all the five quadrupole degrees of 
freedom is a difficult task in many aspects. 
Even if one restricts oneself to 
the rotational degrees of freedom, 
one in general has to deal with integrals over the three Euler angles, 
$\phi$, $\theta$, and $\chi$. 
The full triaxial projection is still too costly, since 
a number of rotated wave functions may be required in order 
to get a converged result. 
Also, the top-GOA scheme for triaxial nuclei 
is not as simple as in the three-level Lipkin model, 
because one has to take into account properly the coupling among the 
three Euler angles. 
How can one overcome these difficulties? 
We shall study here two approximate projection methods.
One is the approximate angular momentum 
projection proposed by Bonche {\it et al.}\cite{BDFH91}, 
which uses the $S_3$ subgroup of the rotation group. 
With this approximation, one needs only five rotated wave 
functions.
The other scheme which we consider is the axially symmetric approximation, 
where the energy is minimized with respect to the deformation $\beta$ 
only, setting the triaxiality $\gamma$ equal to zero. 
With this approximation, the integrations 
for the $\phi$ and $\chi$ angles become unnecessary, reducing the
projection to a one-dimensional integral over 
$\theta$. 
The axially symmetric approximation has been widely used 
in the mean-field calculations \cite{V73,GRT90}, 
where the approximation seems reasonable given that the most nuclei 
do not have a static triaxial ground state. 
However, it is not obvious whether the approximation remains valid when
the fluctuations around the mean-field configuration are included, 
especially when 
the deformation is small. 

The paper is organized as follows. 
In Sec. II, we set up the model Hamiltonian and 
discuss several approaches. These include the mean-field approximation, 
the full triaxial angular momentum projection and its approximation, 
the axially symmetric approximation, and the top-GOA for the 
axial projection. In Sec. III, we compare these schemes 
with the exact solutions of the Hamltonian obtained from 
the matrix diagonalization. We especially focus on the feasibility of each 
method in realistic systems. 
We then summarize the paper in Sec. IV. 
 
\section{$sd$ boson Hamiltonian}

Consider a $N$ boson system whose Hamiltonian is given by,
\begin{equation}
H=H_0+V=\epsilon\sum_{\mu}d_{\mu}^{\dagger}d_{\mu}
-\frac{1}{2}\sum_\mu Q_{\mu}^{\dagger}Q_{\mu}. 
\label{bosonHam}
\end{equation}
The first term expresses the single-particle Hamiltonian $H_0$, 
while the second 
term is the residual quadrupole-quadrupole interaction. The quadrupole 
operator $Q_\mu$ is defined as 
\begin{equation}
Q_\mu=\lambda_1(s^\dagger\tilde{d}_\mu + d_\mu^\dagger s) 
+ \lambda_2[d^\dagger \tilde{d}]^{(2\mu)},
\end{equation}
where $\tilde{d}_\mu=(-)^\mu\,d_{-\mu}$. 
When $\lambda_1=\lambda_2=0$ and $\epsilon > 0$, 
the ground state is the $s$-boson condensed 
state, whose wave function is given by $(s^\dagger)^N/\sqrt{N!}\,|\,\rangle$. 
For a finite value of $\lambda_1$ and $\lambda_2$, the Hamiltonian may be 
diagonalized using the number basis given by, 
\begin{equation}
|\{n\}\rangle = |n_s\,n_{d_{-2}}\,n_{d_{-1}}\,n_{d_0}\,n_{d_1}\,n_{d_2}\rangle,
\end{equation}
taking only the configurations satisfying
\begin{eqnarray}
n_s+n_{d_{-2}}+n_{d_{-1}}+n_{d_{0}}+n_{d_{1}}+n_{d_{2}}&=&N, \label{number}\\
-2n_{d_{-2}}-n_{d_{-1}}+n_{d_{1}}+2n_{d_{2}}&=&0. \label{zproj}
\end{eqnarray}
The first condition, (\ref{number}), constrains the 
boson number, while the second equation, (\ref{zproj}), is the 
condition that the $z$ 
component of the angular momentum is zero. 
With these constraints, the basis has a dimenion of 5 for $N=2$, 18 for $N=4$, 
and 203 for $N=10$. We are going to compare the exact solutions obtained 
in this way with results of the collective treatment based on the mean-field 
approximation plus angular momentum projection. 

\subsection{Mean-field approximation}

We first solve the Hamiltonian in the mean-field approximation. To
this end, we consider 
an intrinsic deformed mean-field state given by \cite{GK80}
\begin{equation}
|\beta\gamma\rangle=\frac{1}{\sqrt{N!}}(b^\dagger)^N\,|\,\rangle,
\label{intrinsic}
\end{equation}
where the deformed boson operator is defined as 
\begin{equation}
b^\dagger=\frac{1}{\sqrt{1+\beta^2}}\left(s^\dagger + \beta\cos\gamma\, 
d_0^\dagger + \frac{\beta}{\sqrt{2}}\sin\gamma\,(d_2^\dagger+d_{-2}^\dagger) 
\right). 
\end{equation}
The parameter $\beta$ accoutns for the global deformation and $\gamma$
for triaxiality.
The deformation energy surface then reads \cite{GK80}
\begin{eqnarray}
E_{\rm MF}(\beta,\gamma)&=&\langle \beta\gamma|H|\beta\gamma\rangle, \\
&=&
\epsilon\,\frac{N\beta^2}{1+\beta^2} \nonumber \\
&&-\frac{1}{2}\,\frac{N}{(1+\beta^2)^2}\,\lambda_1^2
\left\{(1+\beta^2)\left(5+
\left(1+\frac{\lambda_2^2}{\lambda_1^2}\right)\beta^2\right)\right. 
\nonumber \\
&&\left.
+(N-1)\left(4\beta^2-\sqrt{\frac{32}{7}}\,\frac{\lambda_2}{\lambda_1}\,\beta^3
\cos 3\gamma+\frac{2}{7}\,\frac{\lambda_2^2}{\lambda_1^2}\,\beta^4\right)
\right\}. \label{emf}
\end{eqnarray}
One finds that the energy minimum appears on the prolate side 
($\beta > 0, \gamma=0$) when $\lambda_2/\lambda_1 < 0$, while it is 
on the oblate side ($\beta > 0, \gamma=\pi/3$) for $\lambda_2/\lambda_1 > 0$. 
When $\lambda_2$ is zero, the energy surface is independent of $\gamma$, 
corresponding to the $\gamma$-unstable case. 

\subsection{Triaxial Angular Momentum Projection}

When $\beta$ is non-zero, the intrinsic wave function (\ref{intrinsic}) 
is not an eigenstate of the total angular momentum $J$. One can project 
this state onto the $J=0$ state as \cite{RS80}
\begin{equation}
|\beta\gamma,J=0\rangle
\propto \int d\Omega\,\hat{R}(\Omega)\,|\beta\gamma\rangle=
\int^{2\pi}_0d\phi \int^{2\pi}_0d\chi
\int^{\pi}_0\sin\theta d\theta\, \hat{R}(\phi,\theta,\chi)\,
|\beta\gamma\rangle, 
\label{projwf}
\end{equation}
where $\hat{R}(\Omega)$ is the rotation operator. The corresponding 
energy is given by 
\begin{equation}
E_{\rm proj}(\beta,\gamma)
=\frac{\int d\Omega \,\langle\beta\gamma|H\hat{R}(\Omega)|\beta\gamma\rangle}
{\int d\Omega \,\langle\beta\gamma|\hat{R}(\Omega)|\beta\gamma\rangle}. 
\label{eproj}
\end{equation}
Notice that the rotated wave function can be expressed in terms of 
the rotated boson operator as
\begin{equation}
|\beta\gamma\Omega\rangle \equiv \hat{R}(\Omega)\,|\beta\gamma\rangle 
= \frac{1}{\sqrt{N!}}(b_R^\dagger)^N\,|\,\rangle,
\label{rotwf}
\end{equation}
with 
\begin{eqnarray}
b_R^\dagger\equiv\hat{R}(\Omega)b^\dagger\hat{R}^{-1}(\Omega)
&=&
\frac{1}{\sqrt{1+\beta^2}}\left(s^\dagger + \beta\cos\gamma\, 
\sum_mD^2_{m0}(\Omega)d_m^\dagger\right. \nonumber \\
&&\left.+ \frac{\beta}{\sqrt{2}}\sin\gamma\,\sum_m\left(D^2_{m2}(\Omega)
+D^2_{m\,-2}(\Omega)\right)\,d_m^\dagger\right), 
\end{eqnarray}
where $D^2_{mm'}(\Omega)$ is the Wigner's $D$ function. 
The overlaps in the projected energy (\ref{eproj}) can be expressed in
terms of commutators such as
\begin{eqnarray}
[b,b_R^\dagger] &=& \frac{1}{1+\beta^2}\left(
1+\beta^2\cos^2\gamma\,d^2_{00}(\theta)
+\beta^2\sin^2\gamma\,[d^2_{22}(\theta)\cos(2\phi+2\chi)
+d^2_{2-2}(\theta)\cos(2\phi-2\chi)]\right. \nonumber \\
&&\left.\hspace*{1.8cm}+\sqrt{2}\beta^2\sin\gamma\cos\gamma\,
d^2_{20}(\theta)\,[\cos(2\chi)+\cos(2\phi)]\right).
\end{eqnarray}
The results are 
\begin{eqnarray}
I(\Omega)&\equiv&\langle\beta\gamma|\hat{R}(\Omega)|\beta\gamma\rangle =
[b,b_R^\dagger]^N, \\
\frac{H_0(\Omega)}{I(\Omega)}&\equiv&
\frac{
\langle\beta\gamma|H_0\hat{R}(\Omega)|\beta\gamma\rangle}
{\langle\beta\gamma|\hat{R}(\Omega)|\beta\gamma\rangle}
=\epsilon N\left(1-\frac{1}{[b,b_R^\dagger]}\right), \\
\frac{V(\Omega)}{I(\Omega)}&\equiv&
\frac{
\langle\beta\gamma|V\hat{R}(\Omega)|\beta\gamma\rangle}
{\langle\beta\gamma|\hat{R}(\Omega)|\beta\gamma\rangle}
=-\frac{N}{2}\,\frac{1}{[b,b^\dagger_R]}\,
\sum_m\left[[b,Q_m^\dagger],[Q_m,b_R^\dagger]\right] \nonumber \\
&& \hspace*{4.5cm}
-{N(N-1)\over 2 [b,b^\dagger_R]^2}\sum_m\left[b,[Q_m,b_R^\dagger]\right]
\left[[b,Q_m^\dagger],b_R^\dagger\right]. 
\label{Voverlap}
\end{eqnarray}
Here, we have used the relation 
\begin{equation}
[\hat{A},\hat{B}^N] 
=N\hat{B}^{N-1}[\hat{A},\hat{B}]+\frac{1}{2}N(N-1) 
\hat{B}^{N-2}\,[[\hat{A},\hat{B}],\hat{B}]+\cdots, 
\end{equation}
for arbitrary operators $\hat{A}$ and $\hat{B}$. 
We give an explicit expression for the quadrupole commutators, 
$[Q_m,b_R^\dagger]$ and $[Q_m,b^\dagger]$, in the Appendix. 

In practice, one can evaluate the integrals in Eq. (\ref{eproj}) 
as follows. First notice that the integration intervals for the $\chi$ and 
$\phi$ angles can be reduced from $(0,2\pi)$ to $(0,\pi)$ since the $K$ 
quantum number is even for the intrinsic state (\ref{intrinsic}) \cite{BD95}. 
Next, because of the reflection symmetry of the intrinsic wave function 
(\ref{intrinsic}) with respect to the $z$ plane, 
the integration range for the $\theta$ angle can 
be reduced to $(0,\pi/2)$. One can then apply the Gauss-Legendre quadrature 
formula  
to the $\theta$ integral, and the Gauss-Chebyschev formula to the 
$\chi$ and $\phi$ integrals \cite{BD95,ETY99}. 
One may also try the simpler Simpson formula. We will check the convergence 
 of these formulas in the next section. 

\subsection{Approximate Triaxial Projection with Octahedral Group}

Bonche {\it et al.} have considered an approximation to the triaxial 
angular momentum projection (\ref{projwf}) based on the octahedral rotation 
group, that is a group formed from permutations of the principal axes 
of inertia \cite{BDFH91}. With this representation, the 
projected wave function (\ref{projwf}) is approximated as 
\begin{equation}
|\beta\gamma,J=0\rangle\approx
\sum_{i=1}^{24} \hat{S}_i |\beta\gamma\rangle,
\end{equation}
where $\hat{S}_i$ are the 24 elements of the octahedral group. 
In our case with states even under parity, 
the octahedral group is reduced to $S_3$, the group of permutations
of three objects (the $x,y,z$ axes). 
The 6 rotations to be treated are \cite{BDFH91}, 
\begin{eqnarray}
\hat{S}_1&=& \hat{R}(0,0,0)=1, \nonumber \\
\hat{S}_2&=& \hat{R}(\pi,\pi/2,0), \nonumber \\
\hat{S}_3&=& \hat{R}(-\pi/2,-\pi/2,0), \nonumber \\
\hat{S}_4&=& \hat{R}(\pi/2,-\pi/2,\pi/2), \nonumber \\
\hat{S}_5&= &\hat{R}(0,\pi,\pi/2), \nonumber \\
\hat{S}_6&=& \hat{R}(0,\pi,-\pi/2). 
\end{eqnarray}

\subsection{Axial Projection}

When the triaxiality $\gamma$ is zero, the $\phi$ and $\chi$ integrals in 
Eq. (\ref{projwf}) become trivial. The triple integral is then reduced to 
a much simpler single integral with respect to the angle $\theta$. 
This simplifies the projected energy (\ref{eproj}) to
\begin{equation}
E_{\rm proj}(\beta)=\frac{\int^1_{-1}\,d(\cos\theta)\,(H_0(\theta)+V(\theta))}
{\int^1_{-1}\,d(\cos\theta)\,I(\theta)}, 
\label{axial}
\end{equation}
where the overlaps in this axial approximation read 
\begin{eqnarray}
I(\theta)&=&\frac{1}{(1+\beta^2)^N}
\left(1+\frac{\beta^2}{2}(2-3\sin^2\theta)\right)^N, \label{overlap}\\
\frac{H_0(\theta)}{I(\theta)}&=&
\epsilon N\cdot \frac{\beta^2(1-\frac{3}{2}\sin^2\theta)}
{1+\beta^2(1-\frac{3}{2}\sin^2\theta)}, \label{h0overlap}\\
\frac{V(\theta)}{I(\theta)}&=&
-\frac{N}{2}\cdot\frac{1}
{[1+\beta^2(1-\frac{3}{2}\sin^2\theta)]^2}\, \nonumber \\
&&\times\left\{\left(1+\beta^2(1-\frac{3}{2}\sin^2\theta)\right)
\left(5\lambda_1^2+(\lambda_1^2+\lambda_2^2)\beta^2
(1-\frac{3}{2}\sin^2\theta)\right)\right. \nonumber \\
&&\left.+(N-1)\beta^2\left(\lambda_1^2(1+3\cos^2\theta)
+\frac{4}{\sqrt{14}}\lambda_1\lambda_2\beta(1-3\cos^2\theta)\right.\right. 
\nonumber \\
&&
\hspace*{5.5cm}
\left.\left.+\frac{\lambda_2^2}{14}\beta^2(4-9\sin^2\theta\cos^2\theta)
\right)\right\}. \label{voverlap} 
\end{eqnarray}
The axiallay projected energy (\ref{axial}) depends, of course, only
on the global deformation $\beta$. VAP means then to minimize the
projected energy with respect to the deformation parameter $\beta$.

\subsection{Top-GOA for Axial Projection}

A further simplification may be achieved using a second-order approach, 
the extended Gaussian Overlap Approximation (top-GOA). 
In this scheme, the overlaps are expanded up to 
second order derivatives 
with respect to the generator coordinate while retaining 
its topology. For the axial projection considered in the previous subsection, 
the procedure is very similar as in Ref. \cite{HRB02} for the three-level 
Lipkin model. From Eqs. (\ref{overlap} -- \ref{voverlap}), it is clear 
that a natural choice for the expansion variable is $\sin\theta$. Expanding 
the overlaps with respect to $\sin\theta$, one obtains
\begin{eqnarray}
I(\theta)&\approx& \exp\left(-\frac{3}{2}\,\frac{N\beta^2}{1+\beta^2}
\sin^2\theta\right), \\
\frac{H_0(\theta)+V(\theta)}{I(\theta)}&\approx& 
E_{\rm MF}(\beta) + \frac{H_2(\beta)}{2}\sin^2\theta, 
\end{eqnarray}
where $E_{\rm MF}(\beta)$ is the mean-field energy given by 
Eq. (\ref{emf}) (with $\gamma=0$), and $H_2(\beta)$ is defined as 
\begin{equation}
H_2(\beta)=\left.\frac{d^2}{d\theta^2}\,
\frac{H_0(\theta)+V(\theta)}{I(\theta)}\right|_{\theta=0}. 
\end{equation}
Note that we have exponentiated the normalization overlap $I(\theta)$ 
following the idea of the Gaussian overlap approximation \cite{RG87}. 

\section{Numerical Results}

\subsection{Comparison of projection schemes}

The exact ground state for the model Hamiltonian (\ref{bosonHam}) and
the various integrals needed for the projection schemes are solved
numerically by standard methods.
Figure 1 compares the exact solution of the Hamiltonian 
with the several approximations to the triaxial angular 
momentum projection for $N$=4 and $\epsilon=1$. 
The interaction strength $\lambda_2$ 
is set to be $\lambda_2/\lambda_1=-\sqrt{7}/4$ for each $\lambda_1$, 
that is a half the SU(3) value, $(\lambda_2/\lambda_1)_{\rm SU(3)}
=-\sqrt{7}/2$ \cite{E58,A99}.  
The top panel of the figure shows the ground state correlation energy, 
i.e., a difference between the ground state and the mean-field 
energies, as a function of the interaction strength $\lambda_1$. 
The mean-field energy is obtained by minimizing the energy surface 
(\ref{emf}). The optimum deformation parameter $\beta$ thus obtained 
is shown by the thin solid line in the middle panel. 
One sees the phase transition between the spherical and the 
deformed configurations at $\lambda_1=0.47$. The results of full triaxial 
angular momentum projection, obtained by minimizing the projected energy 
surface (\ref{eproj}), are shown by the solid circles in the top panel. These 
results reproduce well the exact results, indicating that the vibrational 
contribution is not large in this model. 
The optimum deformations $\beta$ and $\gamma$ are shown by the thick
solid line in the middle and the bottom panels, respectively. 
In contrast to the mean-field approximation, the optimum deformation $\beta$ 
is finite for all the values of $\lambda_1$, showing no phase 
transition \cite{HRB02}. 
This is a well-known feature of the variation after projection (VAP) 
scheme \cite{RS80}. The dotted line in the figure denotes the results 
of the approximate triaxial angular momentum projection by the 
$S_3$ subgroup of the rotation group. This method does not seem to 
provide enough 
correlation energy, and the agreement with the exact results is poor for 
all the region of $\lambda_1$.  

What is the role played by the triaxiality $\gamma$ in these calculations? 
In order to study this, we show the results of full 
axial projection by the dashed 
line in the figure. 
These are obtained by minimizing the energy function (\ref{axial}),
which is equivalent to minimizing (\ref{eproj}) while keeping $\gamma=$0.
We find that this approximation reproduces the 
exact solution remarkably well. 
The result might appear surprising, since the axially symmetric approximation 
is not expected to work near spherical, where all the five quadrupole 
degrees of freedom should contribute in a similar way. However, as we have 
already discussed, the VAP scheme always leads to a well developed 
deformation even when the mean-field configuration is spherical (see the 
middle panel), and such ``dangerous'' region can be avoided. 
Moreover, even though the 
optimum deformation can be small when the interaction strength 
is very small, this is an irrelevant case since the correlation 
effect is small there. Figure 2 shows the projected energy surface 
$E_{\rm proj}(\beta,\gamma)$, 
measured with respect to the energy 
of the pure configuration, $s^4$, 
at $\lambda_1=0.5$ and $\beta$=0.741 
as a function of triaxiality $\gamma$.  
One sees that the energy gain due to the triaxial deformation is 
indeed small, being consistent with the performance of the 
axially symmetric approximation shown in fig.1. 
We summarize the results for $\lambda_1=0.5$ in Table I. 

As a further test of the axially symmetric approximation, we repeat 
the calculations for $\lambda_2/\lambda_1=0$, that is the $\gamma$ unstable 
case. The results are shown in fig. 3, where the meaning of each line is 
the same as in fig. 1. Note that the optimum triaixiality parameter $\gamma$ 
in the triaxial angular momentum projection is 30 degree for 
all the values of $\lambda_1$, reflecting the $\gamma$ unstable nature of 
the mean-field approximation. In this case, the performance of the axial 
approximation is not as good as in fig.1 (see the dashed line). However, 
it still provides about 80\% of correlation energy at $\lambda_1$ =1, and 
slightly larger at smaller values of $\lambda_1$, which may be acceptable 
even in realistic systems. 

We notice here that the axially symmetric approximation 
is sufficient for $N=2$ irrespective of the value 
of $\lambda_1$ and $\lambda_2$. 
From Eqs. (\ref{projwf}) and (\ref{rotwf}), the (normalized) wave function 
for $J=0$ state reads 
\begin{equation}
|\beta\gamma,J=0\rangle = 
\frac{1}{\sqrt{2+\frac{2}{5}\,\beta^4}}\left.\left.
\left[
\sqrt{2}\,\frac{s^\dagger s^\dagger}{\sqrt{2}}
+\frac{\beta^2}{5}\left(2\,d_2^\dagger d_{-2}^\dagger-2\,
d_1^\dagger d_{-1}^\dagger
+\sqrt{2}\,\frac{d_0^\dagger d_0^\dagger}{\sqrt{2}}\right)\right]\,
\right|\,\right\rangle, 
\label{n=2}
\end{equation}
for any value of $\gamma$.  The projected wave function is thus
independent of $\gamma$, and so is the projected energy surface.  
We also note that the axially symmetric approximation becomes exact in
the limit of $N\to\infty$, as was argued by Kuyucak and Morrison using
the $1/N$ expansion technique \cite{KM88}.  
For $N=2$, the wave
function (\ref{n=2}) is in fact exact, when $\beta$ is minimized.
This follows from the observation there are only two $J=0$ states in
the $(sd)^4$ configuration space, and their relative amplitudes can be
set by a suitable choice of $\beta$, in case of attractive
interactions. 
We have checked the trend in between the two limits, $N=2$ and 
large $N$. The influence of triaxiality are found strongest 
around $N=4$, where the correlation effects are also largest. 
The effect of 
triaxiality then decreases slowly as the boson number $N$ increases. 

\subsection{Efficient angular momentum projection}

We next discuss the feasibility of the angular momentum projection. 
From a computational point of view, it is a costly operation to apply
the rotation operator to a mean-field configuration and take overlaps
with it.  Thus one wants to minimize the number of points in 
the angular integration mesh.  Figure 4 shows the convergence 
of the angular integrals 
in the projected energy surface (\ref{eproj}) with respect to the number 
of rotated wave functions $N_{\rm rot}$, for 
the same parameter set as in fig. 2. 
Notice that the relations $[H,P_J]=0$ and $(P_J)^2=P_J$ are used in deriving 
Eq. (\ref{eproj}), where $P_J$ is the projection operator. 
For a finite value of $N_{\rm rot}$, these relations may be violated, and 
consequently, 
the numerical formula does not give an upper bound of the 
energy. 
The open circles are the results of the Simpson method, while the closed 
circles are obtained with the Gaussian quadrature formulas (see Sec. II-C). 
These are for fixed values of deformation parameters $\beta$ and $\gamma$, 
as indicated in the inset of the figure. The upper panel is for the axial 
projection, while the lower panel for the triaxial projection. Note that 
the former is plotted as a function of $N_{\rm rot}$, while the latter 
involves the three integrals and 
is plotted as a function of $(N_{\rm rot})^{1/3}$. For the Simpson method, 
we exclude the $(\phi,\theta,\chi)=(0,0,0)$ point in counting the number of 
state $N_{\rm rot}$ in the horizontal axis. This state corresponds to 
the unrotated state from which the rotated wave functions are constructed, 
regardless of which quadrature formula one uses. The figure also shows the 
result of top-GOA and the approximate triaxial projection 
with the $S_3$ group as a comparison, which correspond to 
$N_{\rm rot}=1$ and 5, respectively. 
From the figure, one observes that 
the convergence for the axial projection is quick if one uses the 
Gauss-Legendre quadrature formula. The energy is almost converged at 
$N_{\rm rot}$ = 3. The Simpson method, on the other hand, requires 
more terms to achieve the convergence. For the triaxial projection, 
a similar convergence is seen for each of the three integrals. 
However, the required number of rotated wave functions is 
as large as 27 in total, making the triaxial angular momentum projection with 
the VAP minimization impractical. The situation is 
even worse for a larger value of $N$. To demonstrate this, figure 5 shows 
the results for $N=10$. The convergence is somewhat 
slower in this system compared with the $N=4$ case. Note that the 
$N_{\rm rot}$ points-Gauss-Legendre formula is exact when the 
maximum spin in the intrinsic state is 
$J_{\rm max}=2N_{\rm rot}-2$ \cite{BD95,NBT86}. 
In the present $sd$ model, the maximum spin $J_{\rm max}$ is given by 
$2N$, and therefore the more points are needed in order to get a 
convergence for the larger value of $N$. 

Lastly, we discuss the applicability of the top-GOA approach to  
axial projection (see Sec. II-E). This approach requires only one
slightly rotated 
wave function in order to evaluate the second derivatives. 
Figure 6 shows the correlation energy for $N=4$ 
obtained with the top-GOA 
approximation (the dotted line), and with 
the full axial projection (the dashed line). 
The figure also contains the exact solutions as a 
comparison. The upper panel is for $\lambda_2/\lambda_1=-\sqrt{7}/4$, 
while the lower panel is for $\lambda_2/\lambda_1=0$. We see that 
the top-GOA approximation reproduces the full projection reasonably well. 
The performance is somewhat better for $\lambda_2/\lambda_1=-\sqrt{7}/4$. 
As was discussed in Ref. \cite{HRB02}, the applicability of the top-GOA 
approaches increases quickly for a larger value of boson number $N$. 
Indeed, the upper panel of figs. 4 and 5 indicates 
that the agreement between the top-GOA and 
the exact projection significantly improves when $N$=10. 

\section{Summary}

We have used the $sd$-interacting boson model to investigate projections
in a generator coordinate approach to calculate the ground state
correlation energy associated with the quadrupole motion. Our conclusions
about the efficiency of various approximations are clear. The full
angular momentum projection of a triaxial intrinsic state requires 
a large number of rotated wave functions, and it is too costly for
realistic calculations.  On the other hand, we found that the angular momentum
projection of an axial intrinsic state provides a useful ground state
correlation energy. The axially symmetric approximation is exact for
$N=2$ and $N=\infty$. The number of rotated wave
functions needed there is order of 4 if one uses the Gauss-Legendre
quadrature formula to compute the angle integral. The approximate 
triaxial projection using the 
$S_3$ group requires 5 rotated wave functions and still performs
rather 
poorly.  We thus 
conclude that the axial projection provides the most promising 
method to compute systematically the ground state
correlation energy for deformation. 

In applying any projection or generator coordinate expansion, however, 
one has to bear in mind that up to now the energy density functional is 
defined for a single Slater determinant state. It is not designed for
a multi-determinantal wave function such as the projected state, 
and there are ambiguities in calculating the density-dependent interaction 
energy using the energy functional.
Although several recipes have been proposed, they are all subject to 
a conceptional problem. This difficulty can be avoided in either the 
following ways. One is to use the
top-GOA approximation, which can be formulated in terms
of the expectation values in the mean-field wave function
\cite{HRB02}. We have studied the
applicability of the top-GOA with the present model, and have shown
that it already gives a reasonable result for $N=4$ and the
performance improves for larger values of $N$. 
Alternatively, one may also specify the density-dependence in more
detail to remove ambiguities.  Along these lines, a new form of the Skyrme 
interaction was recently proposed by Duguet and Bonche \cite{DB03}.
In either way, the axially symmetric approximation 
leads to a substantial simplification to perform the angular momentum 
projection with only a few Slater determinants, providing 
a useful means 
to construct a microscopic global theory
for the nuclear binding energy systematics.

\section*{acknowledgments}

The authors wish to thank 
H. Flocard, P.-H. Heenen, E. Khan, J. Libert, P. Schuck, Nguyen Van Giai, 
and N. Vin Mauh,
for discussions motivating this study. K.H. and G.F.B. also thank the
IPN Orsay for their warm hospitality where this work was carried out.
G.F.B. also received support from the Guggenheim Foundation and the
U.S. Department of Energy.  K.H. acknowledges support from the
the Kyoto University Foundation. 
P.-G.R. acknowledges support from the
Bun\-des\-ministerium f\"ur Bildung und Forschung (BMBF),
Project No.\ 06 ER 808.

\begin{appendix}
\section{Quadrupole commutators}

In this Appendix, we give an explicit expression for the quadrupole 
commutators, $[Q_m,b_R^\dagger]$ and $[Q_m,b^\dagger]$, in 
Eq. (\ref{Voverlap}). For this purpose, it is convenient to use a compact 
notation for the boson operator, $b_{lm}$, where $b_{00}=s$ and 
$b_{2m}=d_m$. Using this notation, we express the quadrupole operator 
$Q_m$ and the rotated boson operator $b_R^\dagger$ as 
\begin{equation}
Q_m=
\sum_{l_1,m_1}\sum_{l_2,m_2}q^{(m)}_{l_1m_1,l_2m_2}
b_{l_1m_1}^\dagger b_{l_2m_2},
\label{q}
\end{equation}
and
\begin{equation}
b_R^\dagger 
=\sum_{l,m}B_{lm}(\Omega)\,b_{lm}^\dagger,
\label{br}
\end{equation}
respectively. Here, the coefficients $q^{(m)}_{l_1m_1,l_2m_2}$ and 
$B_{lm}(\Omega)$ are given by,
\begin{eqnarray}
q^{(m)}_{00,2m_1}&=&(-)^m\lambda_1\,\delta_{m_1,-m}, \\
q^{(m)}_{2m_1,00}&=&\lambda_1\,\delta_{m_1,m}, \\
q^{(m)}_{2m_1,2m_2}&=&(-)^{m_2}\langle 2\,m_1\,2\,-m_2|2\,m\rangle\,
\lambda_2, \\
B_{00}(\Omega)&=&\frac{1}{\sqrt{1+\beta^2}}\, , \\
B_{2m}(\Omega)&=&
\frac{1}{\sqrt{1+\beta^2}}
\left(\beta\cos\gamma\,D^2_{m0}(\Omega)+\frac{\beta}{\sqrt{2}}
\sin\gamma\,\left(D^2_{m2}(\Omega)
+D^2_{m\,-2}(\Omega)\right)\right). 
\end{eqnarray}
From Eqs. (\ref{q}) and (\ref{br}), one finds 
\begin{equation}
[Q_m,b_R^\dagger] = 
\sum_{l_1,m_1}\sum_{l_2,m_2}q^{(m)}_{l_1m_1,l_2m_2}B_{l_2m_2}(\Omega)
\,b_{l_1m_1}^\dagger. 
\label{qcommu}
\end{equation}
The commutator $[Q_m,b^\dagger]$ can be obtained by setting $\Omega=0$ in 
Eq. (\ref{qcommu}). This yields, 
\begin{eqnarray}
\left[[b,Q_m^\dagger],[Q_m,b_R^\dagger]\right] 
&=&
\sum_{l_1,m_1}\sum_{l_2,m_2}\sum_{l_3,m_3}
q^{(m)}_{l_1m_1,l_2m_2}q^{(m)}_{l_1m_1,l_3m_3}
B_{l_2m_2}(\Omega)B_{l_3m_3}(0), \\
\left[b,[Q_m,b_R^\dagger]\right]
&=&
\sum_{l_1,m_1}\sum_{l_2,m_2}
q^{(m)}_{l_1m_1,l_2m_2}B_{l_2m_2}(\Omega)B_{l_1m_1}(0), \\
\left[[b,Q_m^\dagger],b_R^\dagger\right]
&=&
\sum_{l_1,m_1}\sum_{l_2,m_2}
q^{(m)}_{l_1m_1,l_2m_2}B_{l_1m_1}(\Omega)B_{l_2m_2}(0). 
\end{eqnarray}

\end{appendix}

\begin{table}[hbt]
\caption{ Comparison of 
the ground state energy $E$ and the optimum deformation parameters $\beta$ 
and $\gamma$ obtained with several methods. The parameters of the 
Hamiltonian are taken to be $N=4$, $\epsilon=1$, $\lambda_1=0.5$, and 
$\lambda_2/\lambda_1=-\sqrt{7}/4$.  The energy is measured with respect to
that of the pure configuration, $s^4$. 
}

\begin{center}
\begin{tabular}{c|c|c|c}
Scheme & $E-E(s^4)$ & $\beta$ & $\gamma$ (deg.) \\
\hline
Exact  & $-$0.8193 & --  & -- \\
Triaxial Projection  & $-$0.8189 & 0.741 & 17.64\\
Axial Projection  & $-$0.8017 & 0.723 & 0.0\\
\end{tabular}
\end{center}
\end{table}
\begin{figure}
  \begin{center}
    \leavevmode
    \parbox{0.9\textwidth}
           {\psfig{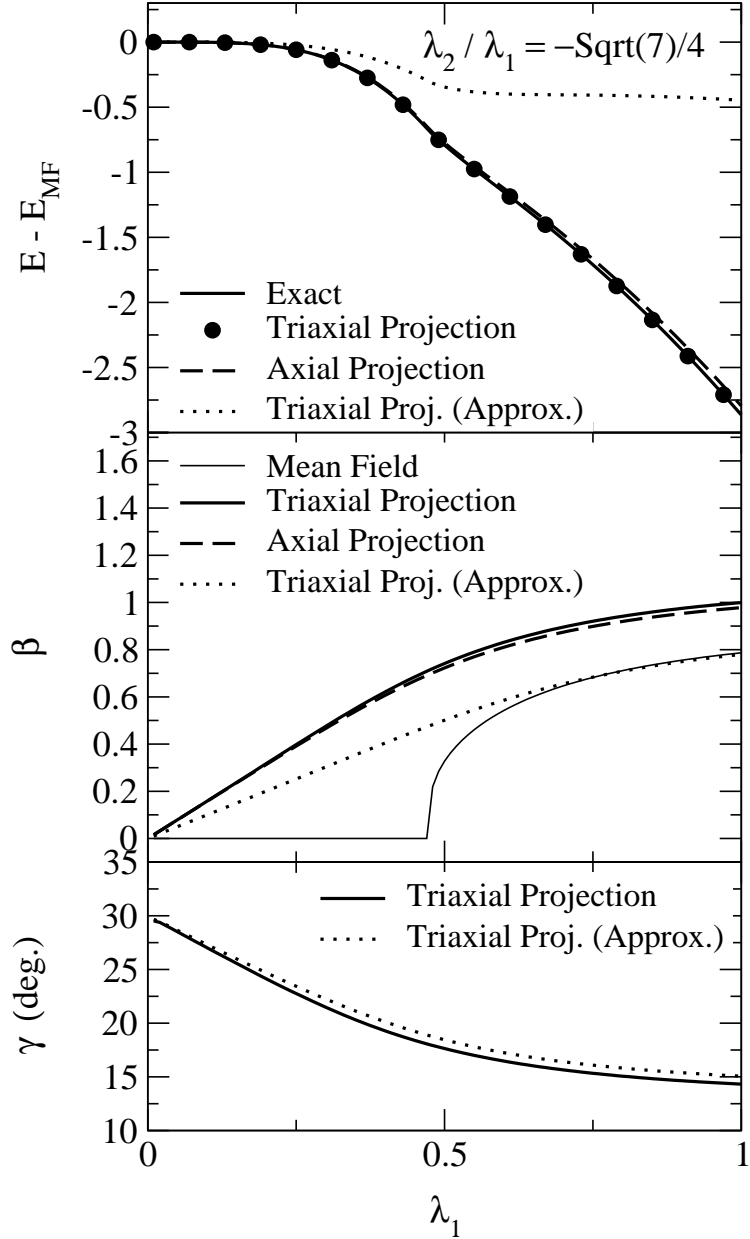}}
  \end{center}
\protect\caption{
The ground state correlation energy obtained by the several methods
    (the top panel). The parameters of the Hamiltonian 
are taken 
to be $N=4, \epsilon=1$, and $\lambda_2/\lambda_1=-\sqrt{7}/4$. 
The solid line is the exact solution of the Hamiltonian obtained by
    the matrix diagonalization. The dots are the results of the full
    triaxial angular momentum projection, while the dashed line is
    obtained by restricting the intrinsic state 
to the axially symmetric shape in minimizing 
the projected energy surface. The dotted line denotes the results of
    the approximate triaxial angular momentum projection which uses
    the $S_3$ subgroup of the octahedral group. 
The middle and the bottom panels show the optimum value of the
    deformation parameters, $\beta$ and $\gamma$, for the angular momentum 
projections. The
    meaning of the thick solid, the dashed, and the dotted lines is
    the same as in the top panel, while the thin solid line is the
    result of the mean-field approximation. }
\end{figure}

\vspace*{3cm}

\begin{figure}
  \begin{center}
    \leavevmode
    \parbox{0.9\textwidth}
           {\psfig{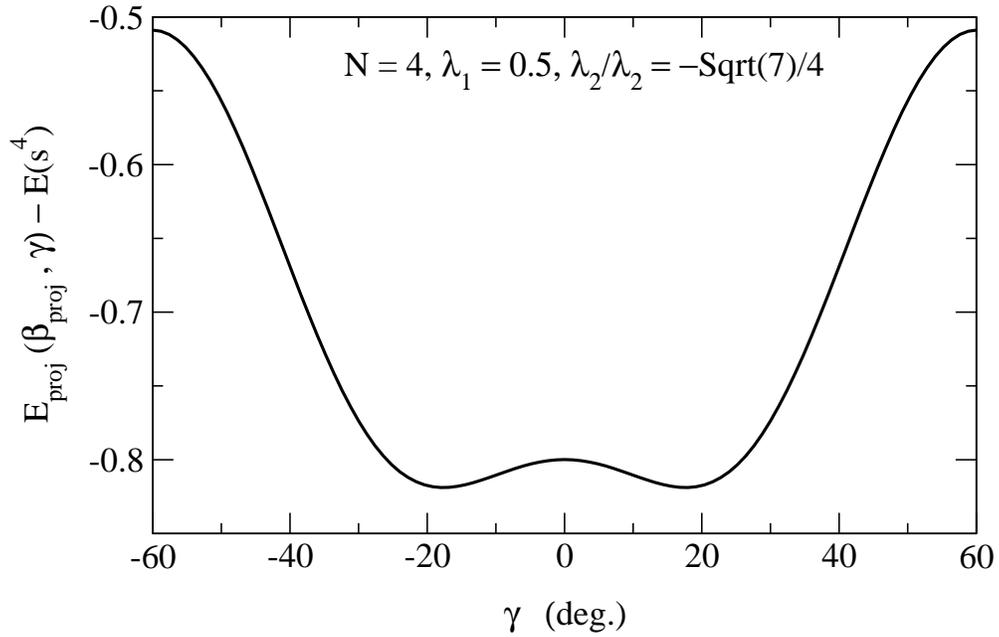}}
  \end{center}
\protect\caption{
The projected energy surface $E_{\rm proj}(\beta,\gamma)$, 
measured with respect to the energy 
of the pure configuration, $s^4$, 
along the $\gamma$
    direction for $\beta=0.741$. 
The parameters of the Hamiltonian 
are taken 
to be $N=4, \epsilon=1$, $\lambda_1=0.5$, and 
$\lambda_2/\lambda_1=-\sqrt{7}/4$. }
\end{figure}

\begin{figure}
  \begin{center}
    \leavevmode
    \parbox{0.9\textwidth}
           {\psfig{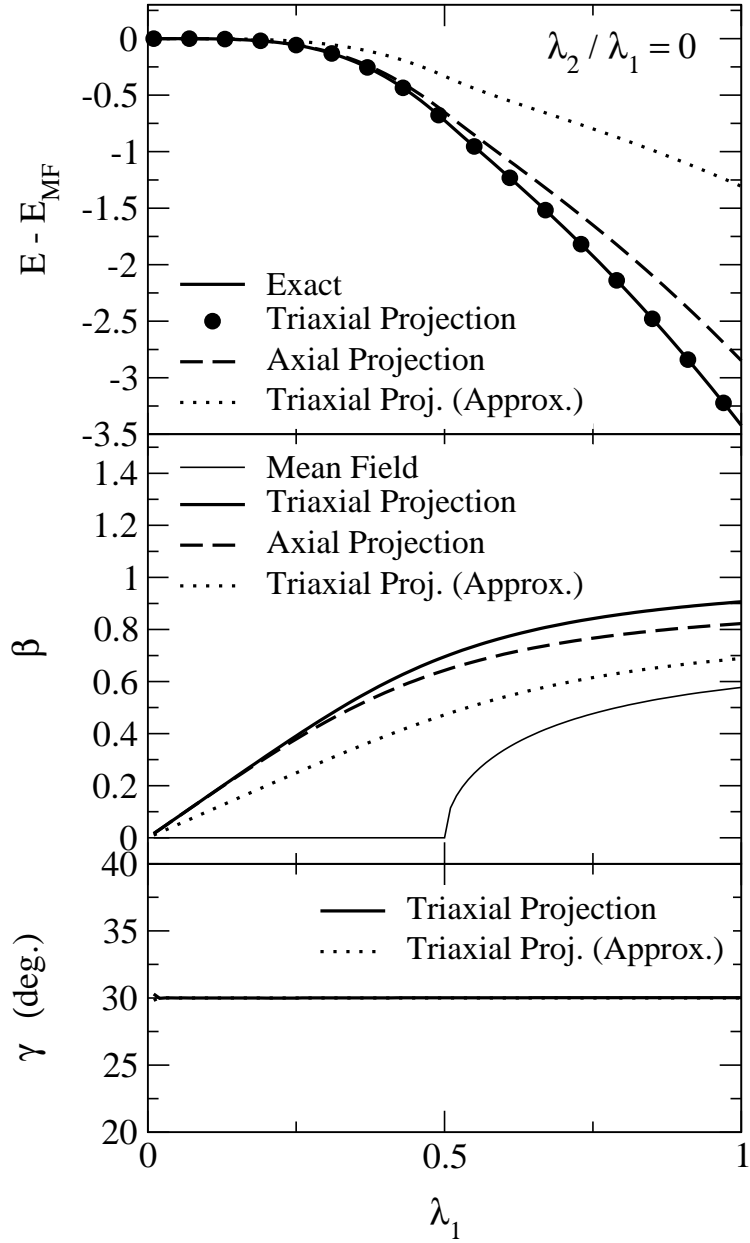}}
  \end{center}
\protect\caption{
Same as fig.1, but for $\lambda_2/\lambda_1=0$. }
\end{figure}

\begin{figure}
  \begin{center}
    \leavevmode
    \parbox{0.9\textwidth}
           {\psfig{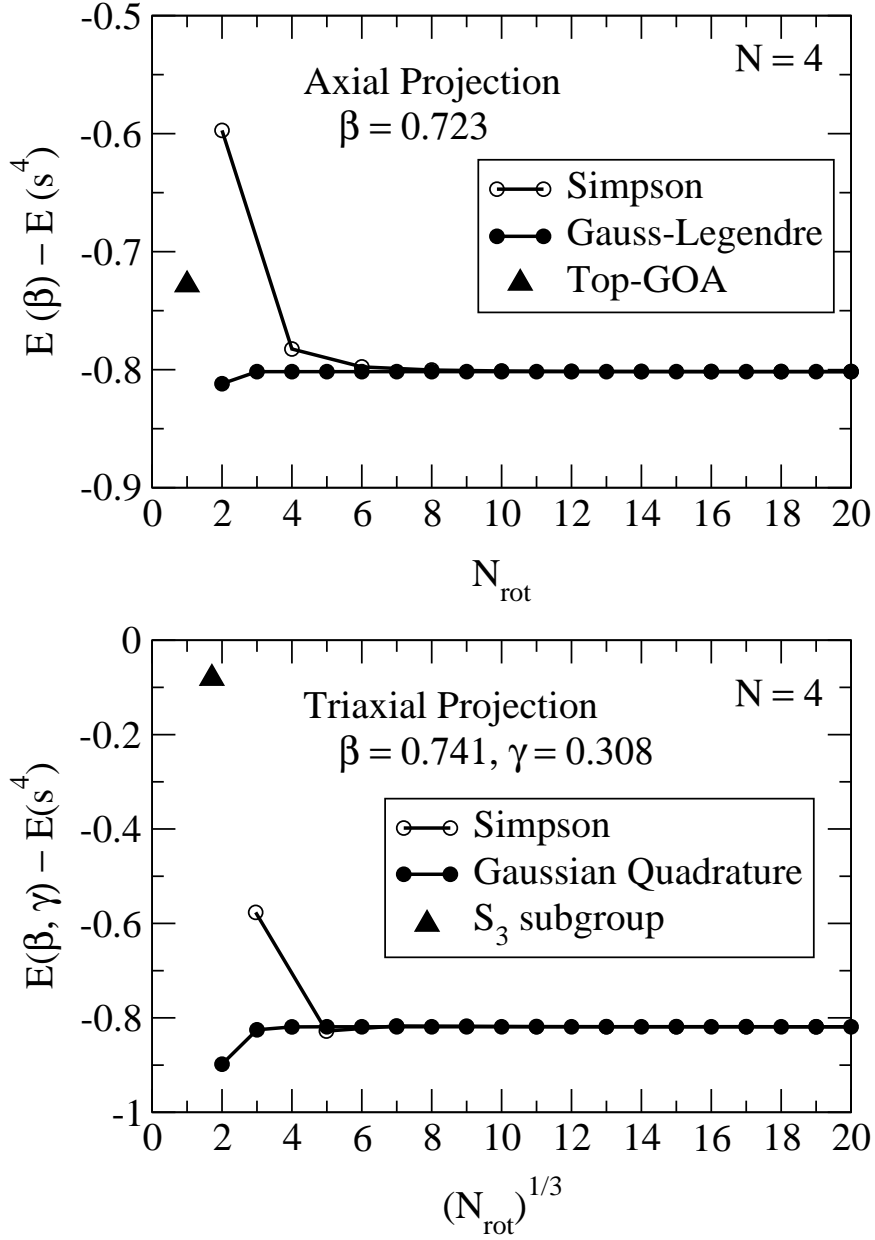}}
  \end{center}
\protect\caption{
Influence of the generator coordinate truncation on the ground state 
energy. The upper and the lower panels are for the axial and the
    triaxial projections, respectively. 
The former plots the energy as a function of the number of 
rotated wave functions $N_{\rm rot}$, while the latter plots as a
    function of $(N_{\rm rot})^{1/3}$, 
for the optimum values of the
    deformation parameters $\beta$ and $\gamma$ indicated in the
    insets. The open and the closed
    circles are the results of the Simpson method and the Gaussian
    quadrature formula, respectively. The triangles denote the result of the
    top-GOA approximation (in the upper panel) and that of the
    approximate projection with the $S_3$ subgroup (in the lower panel). 
The parameters of the Hamiltonian are the same as in fig. 2. }
\end{figure}

\begin{figure}
  \begin{center}
    \leavevmode
    \parbox{0.9\textwidth}
           {\psfig{file=fig5.eps,width=0.7\textwidth}}
  \end{center}
\protect\caption{
Same as fig.4, but for $N=10$. }
\end{figure}

\begin{figure}
  \begin{center}
    \leavevmode
    \parbox{0.9\textwidth}
           {\psfig{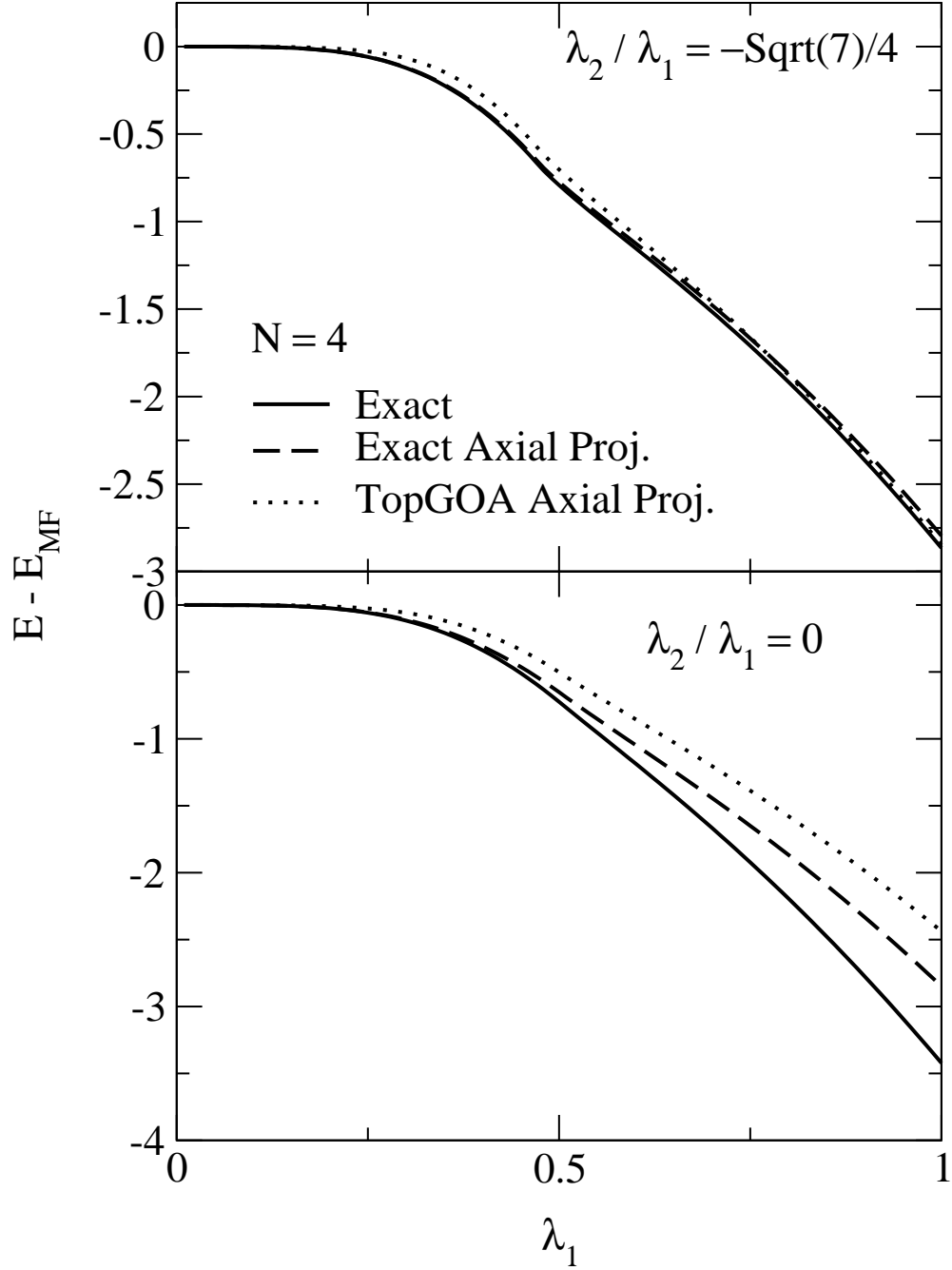}}
  \end{center}
\protect\caption{
The correlation energy obtained in the axially symmetric approximation 
as a function of $\lambda_1$ for $N=4$. The upper and the lower panels
    are for $\lambda_2/\lambda_1=-\sqrt{7}/4$ and $\lambda_2=0$,
    respectively. The dashed line is the result of the full axial
    projection, while the dotted line is obtained in 
the top-GOA approximation to
    the axial projection. The exact solution of the Hamiltonian is
    denoted by the solid line. }
\end{figure}

\end{document}